\documentclass[a4paper,11pt]{article}
\usepackage{float}
\usepackage{jheppub}
\usepackage{cancel}
\usepackage{amsmath}
\usepackage{amsfonts}
\usepackage{enumitem}
\usepackage[numbers]{natbib}
\usepackage[titletoc,toc,title]{appendix}
\usepackage{epstopdf}
\DeclareGraphicsExtensions{.eps}
\usepackage{braket}
\usepackage{tikz}
\usepackage{hyperref}
\hypersetup{
    colorlinks=true,
    linkcolor=blue,
    filecolor=red,      
    urlcolor=blue,
    citecolor=red
} 
\usepackage{braket}
\usepackage{tikz}
\usepackage[bottom]{footmisc}

\title{Holographic Quantum Entanglement Negativity }

\author[a,b]{Pankaj Chaturvedi,}
\author[a]{Vinay Malvimat}
\author[a]{and Gautam Sengupta}

\affiliation[a]{
Department of Physics,\\Indian Institute of Technology Kanpur,\\208016, India}
\emailAdd{cpankaj1@gmail.com}
\emailAdd{vinaymm@iitk.ac.in}
\emailAdd{sengupta@iitk.ac.in}
\affiliation[b]{ Yau Mathematical Sciences Center, Tsinghua University, Beijing, China}
\abstract{\noindent
We advance a holographic conjecture for the entanglement negativity of bipartite quantum states in $(1+1)$-dimensional conformal field theories in the $AdS_3/CFT_2$ framework. Our conjecture exactly reproduces the replica technique results in the large central charge limit, for both the pure state described by the $CFT_{1+1}$ vacuum dual to bulk the pure $AdS_3$ geometry and the finite temperature mixed state dual to a Euclidean BTZ black hole respectively. The holographic entanglement negativity characterizes the distillable entanglement and reduces to a specific sum of holographic mutual informations. We briefly allude to a possible higher dimensional generalization of our conjecture in a generic $AdS_{d+1}/CFT_{d}$ scenario.
}

\begin{document}
\maketitle
\tableofcontents
\pagebreak


\section{Introduction}


Over the last decade quantum entanglement has emerged as a significant theme in diverse disciplines from condensed matter physics to quantum gravity and has attracted focused research attention. In this context entanglement entropy has emerged as 
a crucial measure to characterize entanglement for bipartite pure states. In quantum information theory the entanglement entropy is defined as the von Neumann entropy of the reduced density matrix. Although this quantity is relatively simple to compute for systems with finite degrees of freedom, it is a complex issue for extended quantum many body systems. However 
this issue was addressed in \cite{Calabrese:2004eu} where the authors utilized a replica technique to compute the entanglement entropy for a bipartite system described by a $(1+1)$-dimensional conformal field theory $(CFT_{1+1})$.

Recently there has been a surge of interest in studying the entanglement entropy of holographic $CFT$s in the context of the $AdS/CFT$ correspondence which was inspired through an elegant conjecture advanced by Ryu and Takayanagi in \cite {Ryu:2006bv,Ryu:2006ef}. Their conjecture states that the entanglement entropy 
$S_A$ for a subsystem $A$ in a $(d)$-dimensional holographic CFT ($CFT_d$) is proportional to the area of the co-dimension two bulk extremal surface $\gamma_A$ which is homologous to the subsystem 
$A$ and is given as
\begin{equation}
S_A=\frac{Area(\gamma_A)}{(4G^{(d+1)}_N)},\label{EEI}
\end{equation}
where, $G_N^{(d+1)}$ is the gravitational constant of the bulk $AdS_{d+1}$ space-time. 
Using this holographic prescription it was possible to obtain the entanglement entropy for
bipartite systems described by holographic $CFT_d$s ( see for example the reviews in\cite{ Takayanagi:2012kg, Nishioka:2009un}). 

It is well known however that the entanglement entropy is not a valid measure for the characterization of mixed state entanglement in quantum information theory. For example
the entanglement entropy for a finite temperature mixed state
of a holographic $CFT$ receives contributions from both thermal and quantum correlations \cite{Fischler:2012ca,Chaturvedi:2016kbk}. The characterization of mixed state entanglement was hence a subtle and complex issue in quantum information theory which was addressed in a classic communication by Vidal and Werner \cite {PhysRevA.65.032314}. The authors in this work proposed a computable measure termed {\it entanglement negativity} which characterizes mixed state entanglement for bipartite systems and provides an upper bound to the {\it distillable entanglement}. This measure involves a partial transpose of the reduced density matrix $\rho_A$ over one of the subsystems (say $A_2$ ) in a bipartite system  $A=A_1\cup A_2$ \footnote{The computation of entanglement negativity is subtle involving a procedure termed {\it purification} in quantum information theory. This requires embedding the given bipartite system in a mixed state into a larger auxiliary
system such that the full tripartite system is in a pure state. The auxiliary system may then be traced over to obtain the corresponding reduced density matrix for the bipartite system.}. The monotonicity and non convexity properties of entanglement negativity were later demonstrated in an important communication by Plenio in \cite {Plenio:2005cwa}. 

Interestingly in \cite{Calabrese:2012ew, Calabrese:2012nk, Calabrese:2014yza} the authors provided a replica technique to compute the entanglement negativity for bipartite quantum states
of a $CFT_{1+1}$. Typically for the bipartite finite temperature mixed states of a $CFT_{1+1}$, the authors were able to demonstrate that the entanglement negativity which characterizes the upper bound on the distillable entanglement, involves the elimination of the thermal contributions \cite {Calabrese:2014yza}. Furthermore it could also be shown that for the pure state described by a $CFT_{1+1}$ vacuum, the entanglement negativity reduces to the Renyi entropy of order half as expected from quantum information theory. 

The above discussion naturally leads to the crucial issue of a possible holographic description of the entanglement negativity for bipartite systems described by holographic $CFT$s in the $AdS/CFT$ framework. In the recent past several attempts has been made to understand this critical issue. In this context, the authors in \cite{Rangamani:2014ywa} have computed the entanglement negativity for bipartite systems in pure states described by 
the $CFT$ vacuum which is dual to bulk pure $AdS$ space time. Moreover, in  \cite{Banerjee:2015coc} the authors have conjectured a holographic c-function of which the entanglement negativity may be a possible example. Despite this a clear holographic prescription for the entanglement negativity of bipartite states described by $CFT_d$s in the $AdS/CFT$ framework remained a critical open issue.

As a first step towards addressing the significant open issue described above, in this article we advance a
holographic conjecture for the entanglement negativity of bipartite systems described by a $CFT_{1+1}$ in the $AdS_3/CFT_2$ scenario. In this framework it is possible to compare the holographic bulk results with the corresponding replica technique results for the
$CFT_{1+1}$. Hence such an analysis is expected to provide useful insights into the corresponding higher dimensional extension of the holographic negativity conjecture in a generic $AdS_{d+1}/CFT_d$ scenario. 

For the $AdS_3/CFT_2$ scenario our conjecture involves the holographic description of a four point twist correlator related to the entanglement negativity described in \cite {Calabrese:2014yza}, in terms of an algebraic sum of bulk space like geodesics anchored on appropriate intervals relevant to the purification of the corresponding mixed state. Interestingly the holographic negativity following from our conjecture reduces to a sum of holographic mutual informations ( upto a numerical constant), between the intervals relevant to the purification. For the bipartite pure state described by the $CFT_{1+1}$ vacuum which is dual to the bulk pure $AdS_3$ space-time, the holographic entanglement negativity computed through our conjecture reduces to the Renyi entropy of order half as expected from quantum information theory. This exactly matches with the corresponding replica technique results as given in \cite {Calabrese:2012nk}. We then employ our conjecture to compute the entanglement negativity for bipartite finite temperature mixed state dual to a bulk Euclidean BTZ black hole.  Interestingly the holographic entanglement negativity  for the mixed state in question leads to the elimination of the thermal contribution and exactly reproduces the corresponding $CFT_{1+1}$ results obtained through the replica technique \cite {Calabrese:2014yza}, in the large central charge limit. 

The $AdS_3/CFT_2$ results described above indicate a possible higher dimensional extension of our conjecture relevant to
the entanglement negativity for bipartite states described by  holographic $CFT_d$s in the $AdS_{d+1}/CFT_d$ scenario.  
Such a conjecture is expected to have significant implications for applications to entanglement issues in diverse fields like condensed matter physics, quantum information and issues of quantum gravity.

\section{ Entanglement negativity in CFT$_{1+1}$} As discussed in the introduction, entanglement negativity
leads to the characterization of entanglement for bipartite mixed states in quantum information theory  by providing an upper bound on the distillable entanglement. A suitable replica technique was developed by Calabrese et al. in \cite{Calabrese:2012ew, Calabrese:2012nk,Calabrese:2014yza} for pure and mixed state configurations in $CFT_{1+1}$ which we proceed to briefly describe. For this purpose we consider the tripartition involving the spatial intervals denoted as $A_1$,$A_2$ and $B$ such that $A_1$ and $A_2$ correspond to finite intervals defined by $[u_1,v_1]$ and $[u_2,v_2]$ of lengths $l_1$ and $l_2$ respectively whereas $B$ represents the rest of the system. The reduced density matrix $\rho_A$ of the subsystem $A=A_1\cup A_2$ is obtained by tracing over the subsystem $B$ i.e. $\rho_A=Tr_B\rho$. The definition of entanglement negativity requires the operation of partial transpose over one of the subsystem. If $|q_i^1\big>$ and $|q_i^2\big>$ represent the bases of Hilbert space corresponding to subsystems  $A_1$ and $A_2$ respectively then this operation of the partial transpose with respect to $A_2$ degrees of freedom is defined as follows
\begin{equation}
\big<q_i^1q_j^2|\rho_{A}^{T_{2}}|q_k^1q_l^2\big> = \big<q_i^1q_l^2|\rho_{A}|q_k^1q_j^2\big>,\label{ptrace}
\end{equation}
The authors in \cite{Calabrese:2012ew, Calabrese:2012nk} provided a replica definition for the entanglement negativity as\footnote{Note that this definition reduces to the one employed in quantum information theory ${\cal E}=\ln||\rho_A^{T_2}||$ provided by Vidal and Werner, in the replica limit $n_e\to 1$.}
\begin{equation}
{\cal E}=\lim_{n_e \to 1}\ln[Tr(\rho_A^{T_2})^{n_e}],\label{ENCFT}
\end{equation}
where $n_e$ denotes that the parity of $n$ is even. This is because the quantity $Tr(\rho_A^{T_2})^{n}$ shows different functional dependence on the eigenvalues of $\rho_A^{T_2}$  based on the parity of $n$. A sensible result is obtained by defining entanglement negativity to be an analytic continuation of even sequences of $n_e$ to $n_e=1$ as suggested in \cite{Calabrese:2012ew, Calabrese:2012nk} \footnote{ \label{Note17} The explicit construction of this non-trivial analytic continuation remains elusive except for some simple conformal field theories ( See \cite{Calabrese:2009ez,Calabrese:2010he,Calabrese:2014yza} ).}.  In a $CFT_{1+1}$ the operation of partial transpose on the reduced density matrix $\rho_A$ under the trace i.e $Tr(\rho_A^{T_2})$ has the effect of exchanging upper and lower edges of the branch cut along the interval $A_2$ on the $n_e$-sheeted Riemann surface. This leads to the following expression for the quantity $Tr(\rho_A^{T_2})^{n_e}$ in terms of the twist field correlator,
\begin{equation}
Tr(\rho_A^{T_2})^{n_e}=\big<{\cal T}_{n_e}(u_1)\overline{{\cal T}}_{n_e}(v_1)\overline{{\cal T}}_{n_e}(u_2){\cal T}_{n_e}(v_2)\big>.\label{rhoAn3}
\end{equation}
In the limit $v_1\rightarrow u_2$ and $v_2\rightarrow u_1$, the tripartite configuration $(A_1,A_2,B)$ reduces to a bipartite configuration $(A,B,\emptyset)$ such that the interval corresponding to the subsystem $A$ is now given by $[u,v]$ with its length given as $l=|u-v|$. Hence, in this limit the quantity
$Tr(\rho_A^{T_2})^{n_e}\to Tr(\rho^{T_{A^c}})^{n_e}= Tr(\rho^{T_A})^{n_e}$ and the four point function described above reduces to the following product of two point twist correlators
\begin{equation}
Tr(\rho^{T_A})^{n_e}=\big<{\cal T}^2_{n_e}(u)\overline{{\cal T}}^2_{n_e}(v)\big>.\label{rhoAnzero}
\end{equation}
The authors in \cite{Calabrese:2012ew,Calabrese:2012nk} observed that the twist fields ${{\cal T}}^2_{n_e}$ connect $n_e^{th}$ sheet of the Riemann surface to $(n_e+2)^{th}$ sheet of the Riemann surface. Similarly, the twist field $\overline{{\cal T}}^2_{n_e}$ connects $n_e^{th}$ sheet to $(n_e-2)^{th}$ sheet of the Riemann surface. This leads to the following factorization of the correlation function in eq.(\ref{rhoAnzero}) due to the 
reduction of the $n_e$ ($n$ even) sheeted Riemann surface into two $n_e/2$ sheeted Riemann surfaces,  
\begin{equation}
 \langle{\cal T}^2_{n_e}(u)\overline{{\cal T}}^2_{n_e}(v)\rangle_{\mathbb{C}}=\langle{\cal T}_{\frac{n_e}{2}}(u)\overline{{\cal T}}_{\frac{n_e}{2}}(v)\rangle_{\mathbb{C}}^2.\label{42f}
\end{equation}
 Hence, the scaling dimension $(\Delta_{n_e}^{(2)})$ of the operator ${\cal T}_{n_e}^2$ can be related to the scaling dimensions $(\Delta_{n_e})$ of the operator ${\cal T}_{n_e}$  as 
\begin{eqnarray}
\Delta_{n_e}^{(2)}&=&2\Delta_{\frac{n_e}{2}}= \frac{c}{6}\left(\frac{n_e}{2}-\frac{2}{n_e}\right),\\\Delta_{n_e}&=& \frac{c}{12}\left(n_e-\frac{1}{n_e}\right).\label{dimTn2}
\end{eqnarray}
This leads to the following expression for the entanglement negativity of a bipartite pure state described by the vacuum of the $CFT_{1+1}$ as
\begin{equation}
{\cal E}=\frac{c}{2}\ln\left(\frac{l}{a}\right)+constant=\frac{3}{2}S_A+constant,\label{ENCFTzero}
\end{equation}
where, $a$ is the UV cut-off for the $CFT_{1+1}$, $S_A$ denotes the entanglement entropy of the subsystem
$A$ in the pure vacuum state of the $CFT_{1+1}$ and the non-universal constant is related to the normalization of the two point twist field correlator. This result conforms to quantum information expectation that the entanglement negativity for a bipartite pure state is given by the Renyi entropy of order half. However, in a subsequent significant communication \cite{Calabrese:2014yza} the authors demonstrated through explicit geometrical arguments that although the result in eq.(\ref{rhoAnzero}) is valid for a bipartite pure  state of the $CFT_{1+1}$ it is incorrect for the finite temperature mixed state of the $CFT_{1+1}$ \footnote{This is because
the quantity $Tr(\rho^{T_A})^{n_e}$ does not factorize in the finite temperature case due to subtle geometric reasons. See \cite{Calabrese:2014yza} for a detailed explanation of this.}. 
It could then be shown that the correct expression for the entanglement negativity of the bipartite finite temperature mixed state is given in terms of a four point twist field correlator as follows 
\begin{equation}
{\cal E}=\lim_{L \to \infty}\lim_{n_e \to 1}\ln[\big<{\cal T}_{n_e}(-L)\overline{{\cal T}}^2_{n_e}(-l){\cal T}^2_{n_e}(0)\overline{{\cal T}}_{n_e}(L)\big>_{\beta}], \label{ENCFT1}
\end{equation}
where ${\cal T}_{n}(L)$ corresponds to the twist field at some large distance $L$ from the interval $A$. It is to be noted that the order of limits plays a crucial role for this case implying that the bipartite limit $(L \rightarrow \infty)$ should be taken only after the replica limit $(n_e\rightarrow 1)$. Note that the subscript $\beta$ in eq.(\ref{ENCFT1}) indicates
that the four point twist correlator has to be evaluated on an infinitely long cylinder of circumference $\beta=1/T$. This cylindrical geometry can be obtained from the complex plane by the conformal transformation $z\rightarrow \omega=\beta/2 \pi\ln z$ where $z$ denotes the coordinates on the complex plane and $\omega$ denotes the coordinates on the cylinder. The form of the four point function on the complex plane is fixed upto a function of cross ratios as follows
\begin{eqnarray}
\big< {\cal T}_{n_e}(z_1)\overline{{\cal T}}^2_{n_e}(z_2){\cal T}^2_{n_e}(z_3)\overline{{\cal T}}_{n_e}(z_4)\big>_{\mathbb{C}}\nonumber=\frac{c_{n_e}c^2_{n_e/2}}{z_{14}^{2\Delta_{n_e}}z_{23}^{2\Delta^{(2)}_{n_e}}}\frac{{\cal F}_{n_e}(x)}{x^{\Delta^{(2)}_{n_e}}},\\\label{rhoAn5}
\end{eqnarray}
where the cross ratio $x=\frac{z_{12}z_{34}}{z_{13}z_{24}}$,$z_{ij}=|z_i-z_j|$,  $c_{n_e}$ and $c^2_{n_e/2}$ are the normalization constants of the two point twist correlator. The subscript ${\mathbb{C}}$ in the above equation denotes 
that the four point twist correlator is evaluated on the complex plane. Following \cite{Calabrese:2014yza} it is possible to obtain the following constraints on the arbitrary function ${\cal F}_{n_e}(x)$ in the two limits $x\to 1$ and $x \to 0$,
\begin{equation}
{\cal F}_{n_e}(1)=1,~~~{\cal F}_{n_e}(0)=C_{n_e},\label{cons1}
\end{equation}
here, $C_{n_e}$ is a constant that depends on the full operator content of the theory. The entanglement negativity for the bipartite finite temperature mixed state of the $CFT_{1+1}$ may then be obtained from eq.(\ref{ENCFT1}) and eq.(\ref{rhoAn5}) through a conformal map from the complex plane to the cylinder. This leads to the following expression for the entanglement negativity\cite{Calabrese:2014yza}
\begin{eqnarray}
{\cal E}&=&\frac{c}{2}\ln\left[\frac{\beta}{\pi a}\sinh\left(\frac{\pi l}{\beta}\right)\right]-\frac{\pi c l}{2\beta}+f(e^{-2\pi l/\beta})+const.\nonumber\\ \nonumber \\
 \mathrm{where}&~~~&f(x)=\lim_{n_e \to 1}\ln[{\cal F}_{n_e}(x)]\nonumber,
\end{eqnarray}
The above expression for the entanglement negativity may be recast into an interesting form as
\begin{equation}\label{ENCFTfinite}
 {\cal E}=\frac{3}{2}\big[S_{A}-S_{A}^{th}\big]+f(e^{-2\pi l/\beta})+const
\end{equation}
where $S_{A}=\frac{c}{3}\ln[\frac{\beta}{\pi a}\sinh\left(\frac{\pi l}{\beta}\right)]$ and $S_{A}^{th}=\frac{\pi c l}{3\beta}$ correspond to the entanglement entropy and the thermal entropy of the subsystem $A$ for the finite temperature mixed state respectively. This result clearly suggests that the entanglement negativity precisely leads to the upper bound on the distillable entanglement for the finite temperature mixed state through the elimination of the thermal contribution. Note that the non-universal function $f(e^{-2\pi l/\beta})$ and the constant depend on the full operator content of the theory. 

\section{ Large central charge limit }
In the $AdS_3/CFT_2$ scenario, it is well known that the semiclassical limit in the $AdS_3$ bulk  defined by the limit $G_N\to 0$ corresponds to the large central charge limit $c\to \infty$ in the dual $CFT_{1+1}$, as these two quantities are related by the Brown-Hennaux formula $c=\frac{3R}{2G_{N}^{(3)}}$\cite{brown1986}. Hence, it is required to examine the large central charge limit of the twist field correlator in eq.(\ref{ENCFT1}) for the entanglement negativity before we proceed to a holographic description involving the bulk $AdS_3$ geometry. The large central charge limit of such twist field correlators for the entanglement entropy of multiple intervals were investigated in \cite{Hartman:2013mia,Headrick:2010zt}. The authors in  \cite{Hartman:2013mia} utilized the monodromy technique \cite{Zamolodchikov:1995aa,Harlow:2011ny,Fitzpatrick:2014vua} to determine the large central charge limit of entanglement entropy for multiple disjoint intervals. They demonstrated that to the leading order in $\frac{1}{c}$ the entanglement entropy for multiple intervals  is universal (i.e independent of the full operator content of the theory) and the results match exactly with those computed from the Ryu-Takayanagi conjecture. A different four point twist correlator describing the entanglement negativity of a simpler mixed state configuration of adjacent intervals in $CFT_{1+1}$ has also been studied in the large central charge limit  in \cite{Kulaxizi:2014nma}.  The authors there demonstrated that in this limit the entanglement negativity for the mixed state of adjacent intervals  matches exactly with the universal part obtained through the replica technique by Calabrese et al. in  \cite{Calabrese:2012ew, Calabrese:2012nk}. Notice that in all of the above mentioned cases the dominant contribution to the twist field correlators are universal, whereas the non-universal contributions constitute the sub leading $\frac{1}{c}$ ~corrections in the large central charge limit. 

The above discussion suggests a similar computation to demonstrate that the non-universal function $f(e^{-2\pi l/\beta})$ and the constant in eq.(\ref{ENCFTfinite}) for the entanglement negativity of the mixed state in question, to be sub leading in the large central charge limit ( see  \footnote{We have recently performed this monodromy computation in \cite{Malvimat:2017yaj} for the required four point twist correlator for the entanglement negativity in which we have demonstrated that this non-universal function is indeed sub leading in the large central charge limit.}). 
From \cite{Malvimat:2017yaj} we therefore observe that in the large central charge limit the entanglement negativity in eq.(\ref{ENCFTfinite}) reduces to the following universal expression 
\begin{equation}\label{ENCFTlc}
 {\cal E}=\frac{3}{2}\big[S_{A}-S_{A}^{th}\big].
\end{equation}
Having determined the large central charge limit of the entanglement negativity for the bipartite ($A\cup A^c$) mixed state described by the finite temperature $CFT_{1+1}$, we now proceed to outline our holographic conjecture to describe this quantity through the bulk $AdS_3$ geometry.

\section{Holographic entanglement negativity in AdS$_3$/CFT$_2$}

\begin{figure}[ht!]
\centering
\includegraphics[scale=2.0]{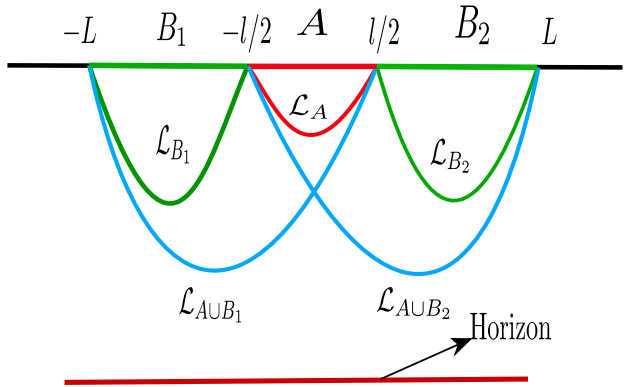}
\caption{\label{fig:subsystem} Schematic of geodesics anchored on the subsystems $A$, $B_1$ and $B_2$ which live on the (1+1)-dimensional boundary.}
\end{figure}

In this section we utilize the standard $AdS/CFT$ dictionary in order to formulate the holographic conjecture for the entanglement negativity in the $AdS_3/CFT_2$ scenario. We begin by first considering a bipartition ( $A\cup A^c$) described by an interval $A$ of length $l$ and its complement $A^c$. For our construction it is also required to consider two other subsystems $B_1$ and $B_2$ within $A^c$ on either side of $A$ as depicted in the fig.(\ref{fig:subsystem}). 

The two point twist correlators are fixed by the conformal symmetry as
\begin{eqnarray}
\big<{\cal T}_{n_e}(z_k)\overline{{\cal T}}_{n_e}(z_l)\big>_{\mathbb{C}}= \frac{c_{n_e}}{z_{kl}^{2\Delta_{n_e}}}~~~~~~~~~~~~~~~\label{2pc2}\\
\langle{\cal T}^2_{n_e}(u)\overline{{\cal T}}^2_{n_e}(v)\rangle_{\mathbb{C}}=\big<{\cal T}_{\frac{n_e}{2}}(z_i)\overline{{\cal T}}_{\frac{n_e}{2}}(z_j)\big>^2_{\mathbb{C}}=\frac{c_{n_e/2}^2}{z_{ij}^{4\Delta_{\frac{n_e}{2}}}},~~~~\label{2pc1}
\end{eqnarray} 
where we have utilized the factorization in eq.(\ref{42f}).
Interestingly, the universal part of the four point twist correlator in eq.(\ref{rhoAn5}), which is dominant in the large central charge limit, factorizes into the above mentioned two point twist correlators as follows
\begin{eqnarray}
  \big< {\cal T}_{n_e}(z_1)\overline{{\cal T}}^2_{n_e}(z_2){\cal T}^2_{n_e}(z_3)\overline{{\cal T}}_{n_e}(z_4)\big>_{\mathbb{C}}=&\big<{\cal T}_{\frac{n_e}{2}}(z_2)\overline{{\cal T}}_{\frac{n_e}{2}}(z_3)\big>^2\big<{\cal T}_{n_e}(z_1)\overline{{\cal T}}_{n_e}(z_4)\big>\nonumber\\&\times\frac{\big<{\cal T}_{\frac{n_e}{2}}(z_1)\overline{{\cal T}}_{\frac{n_e}{2}}(z_2)\big>\big<{\cal T}_{\frac{n_e}{2}}(z_3)\overline{{\cal T}}_{\frac{n_e}{2}}(z_4)\big>}{\big<{\cal T}_{\frac{n_e}{2}}(z_1)\overline{{\cal T}}_{\frac{n_e}{2}}(z_3)\big>\big<{\cal T}_{\frac{n_e}{2}}(z_2)\overline{{\cal T}}_{\frac{n_e}{2}}(z_4)\big>}+O[\frac{1}{c}]\label{factor}
\end{eqnarray}
From the $AdS/CFT$ dictionary it is known that the two point twist correlator in the holographic $CFT_{1+1}$ is  related to the length of the space like geodesic ($L_{ij}$) anchored on the points $(z_i,z_j)$ and extending into the $AdS_{3}$ bulk as follows \cite {Ryu:2006bv,Ryu:2006ef}
\begin{eqnarray}
\big<{\cal T}_{n_e}(z_k)\overline{{\cal T}}_{n_e}(z_l)\big>_{\mathbb{C}} &\sim& e^{-\frac{\Delta_{n_e}{ \cal L}_{kl}}{R}}\label{l34}\\
\big<{\cal T}_{\frac{n_e}{2}}(z_i)\overline{{\cal T}}_{\frac{n_e}{2}}(z_j)\big>_{\mathbb{C}} &\sim& e^{-\frac{\Delta_{\frac{n_e}{2}}{ \cal L}_{ij}}{R}},\label{l12}
\end{eqnarray}
where, $R$ is the $AdS_3$ radius. From fig.(\ref{fig:subsystem}) we identify that 
\begin{eqnarray}
{\cal L}_{12}&=&{\cal L}_{B_1},~~{\cal L}_{23}={\cal L}_{A},~~{\cal L}_{34}={\cal L}_{B_2},\nonumber\\
{\cal L}_{13}&=&{\cal L}_{A\cup B_1},~~ {\cal L}_{24}={\cal L}_{A\cup B_2},~~ {\cal L}_{14}={ \cal L}_{A\cup B}.\label{Lrel}
\end{eqnarray}

Utilizing eq.(\ref{factor}) describing the factorization of the four point twist correlator
into two point correlators, and the geodesic approximations in (\ref{l34}) and (\ref{l12}), it is now possible to express the four point twist correlator in eq.(\ref{rhoAn5}) in a suggestive form as follows
\begin{eqnarray}
\big<{\cal T}_{n_e}(z_1)\overline{{\cal T}}^2_{n_e}(z_2){\cal T}^2_{n_e}(z_3)\overline{{\cal T}}_{n_e}(z_4)\big>_{\mathbb{C}}\sim \exp{[\frac{-\Delta_{n_e} X-\Delta_{\frac{n_e}{2} } Y}{R}]},\label{rhoAn7}
\end{eqnarray}
where
\begin{eqnarray}
X&=&{ \cal L}_{A\cup B} \\Y&= &2{\cal L}_{A}+{\cal L}_{B_1}+{\cal L}_{B_2}-{\cal L}_{A\cup B_1}-{\cal L}_{A\cup B_2}
\end{eqnarray}
Note that the coordinates $(z_1,z_2,z_3,z_4)=(-L,-\frac{l}{2},\frac{l}{2},L)$ in eq.(\ref{rhoAn7}) are as depicted in the fig.(\ref{fig:subsystem}). As discussed earlier, in the large central charge limit we have ignored the sub leading contribution involving the non-universal function $f(x)$ in the above expression \footnote{The authors would like to thank Ashoke Sen for pointing this out.}. In the replica limit $n_e\rightarrow 1$, $\Delta_{n_e}\rightarrow 0$ and $\Delta_{\frac{n_e}{2}}\rightarrow -\frac{c}{8}$ \footnote{Note that the negative scaling dimension of the twist operators (${\cal T}^2_{n_e}$ and ${\cal T}_{\frac{n_e}{2}}$) in the replica limit $n_e\to1$ has to be understood only in the sense of an analytic continuation $n_e\to 1$ (See also footnote \ref{Note17}). }. Hence, the eq.(\ref{rhoAn7}) leads to the following expression for the holographic entanglement negativity 
\begin{equation}
{\cal E}=\lim_{B \to A^c}~\frac{3}{16G_{N}^{(3)}}\bigg[(2 {\cal L}_{A}+{\cal L}_{B_1}+{\cal L}_{B_2}-{\cal L}_{A\cup B_1}-{ \cal L}_{A\cup B_2})\bigg],\label{ENCFT3}
\end{equation}
where we have used the Brown-Hennaux formula $c=\frac{3R}{2G_{N}^{(3)}}$.   
 
Observe that upon utilizing  the Ryu-Takayanagi conjecture in eq.(\ref{EEI}), the above equation eq.(\ref{ENCFT3}) reduces to the following form
 \begin{equation}
{\cal E}=\lim_{B \to A^c}\frac{3}{4}\bigg[2 S_{A}+S_{B_1}+S_{B_2}-S_{A\cup B_1}-S_{A\cup B_2}\bigg],\label{ENCFTSA}
\end{equation}
and  $B\to A^c$ denotes the bipartite limit in which the intervals $B_1, B_2$ are extended to infinity such that $B_1\cup B_2=A^c$. Note that the bipartite limit essentially corresponds to the limit $L \to \infty$ in the $CFT_{1+1}$ described earlier.
Recall that the holographic mutual information between the pair of intervals $(A,B_1)$ and $(A,B_2)$ are given as
\begin{eqnarray}
 {\cal I}(A,B_i)&=& S_{A}+S_{B_i}-S_{A\cup B_i},\nonumber\\
&=&\frac{1}{4G_{N}^{(3)}}({\cal L}_{A}+{ \cal L}_{B_i}-{\cal L}_{A\cup B_i}),~~~~~~\label{MI}
\end{eqnarray}
where $i=\{1,2\}$. Interestingly from  equations (\ref{ENCFTSA}) and (\ref{MI}) we observe that the holographic entanglement negativity may be re expressed in terms of the holographic mutual informations between the intervals
$(A,B_1)$ and $(A,B_2)$ as follows \footnote{Note that entanglement negativity and mutual information are completely distinct measures in quantum information theory. This does not contradict with the expression in eq.(\ref{ENCFT4}) as this result holds only for the universal parts of the two quantities which are dominant in the holographic (large central charge limit $c\to\infty$) limit. Interestingly the matching between the universal parts of these two quantities has also been reported for the case of adjacent intervals in a $CFT_{1+1}$ in connection with both local and global quench \cite{Coser:2014gsa,Wen:2015qwa}.}
\begin{eqnarray}
{\cal E}&=&\lim_{B \to A^c}\frac{3}{4}\big[{\cal I}(A,B_1)+{\cal I}(A,B_2)\big],\label{ENCFT4}
\end{eqnarray}
Clearly the above discussion suggests a holographic conjecture for the entanglement negativity of bipartite systems ($A\cup A^c$) described by holographic $CFT_{1+1}$s through eq. (\ref {ENCFT3}), in terms of the lengths of the bulk space like  geodesics in $AdS_3$ as depicted in the fig.(\ref{fig:subsystem}). 

Observe that
the large central charge analysis described in \cite {Malvimat:2017yaj} for the entanglement negativity in $CFT_{1+1}$ assumes a crucial role in the above arguments leading to our conjecture. We mention here that although the large central charge analysis strongly substantiates our conjecture for the $AdS_3/CFT_2$ scenario, a bulk proof along the lines of \cite {Faulkner:2013yia} remains a non trivial open issue. However as a first consistency check we employ our holographic conjecture described by eq.(\ref{ENCFT3}) to compute the entanglement negativity for specific quantum states of a holographic $CFT_{1+1}$. These include the bipartite pure state described by the $CFT_{1+1}$ vacuum which is dual to the pure $AdS_3$ space time and the finite temperature mixed state dual to the Euclidean BTZ black hole. Quite remarkably we are able to demonstrate that in both the cases the holographic entanglement negativity computed through our conjecture, exactly matches with the corresponding replica technique results in the large central charge limit. We emphasize that the exact reproduction of the $CFT_{1+1}$ results mentioned above is an extremely significant consistency check and provides strong indication towards the plausibility of a formal proof along the lines of \cite {Faulkner:2013yia}. 

\subsection{ Pure AdS$_3$}

According to the AdS/CFT correspondence, the vacuum state of a holographic 
$CFT_{1+1}$ is dual to the pure $AdS_3$ space time whose metric in Poincare coordinates is given as
\begin{equation}
 ds^{2}=\frac{R^2}{z^2}(-dt^2+dz^2+dx^2)\label{puads}.
\end{equation}
where $z$ is the inverse radial coordinate extending into the bulk, $R$ is the AdS radius and $(x,t)$ are the coordinates on the boundary $CFT_{1+1}$.
The length of bulk geodesic ${\cal L}_{\gamma}$  anchored to the subsystem $\gamma$ in the boundary $CFT_{1+1}$ in this space time is given as
\begin{equation}
 {\cal L}_{\gamma}= 2R ~\ln\big[\frac{l_{\gamma}}{a}\big].\label{geopu}
\end{equation}
Utilizing the above expression for the various subsystems $\gamma=\{A,B_1,B_2,A\cup B_1, A\cup B_2\}$ as depicted in the fig(\ref{fig:subsystem}) and substituting these in the expression for the holographic entanglement negativity given by eq.(\ref{ENCFT4}) we obtain
\begin{equation}
 {\cal E}=\frac{3R}{4G_N}\ln\big[\frac{l}{a}\big].\label{henpu}
\end{equation}
where $l$ is the length of the interval $A$ as described earlier and $a$ is the UV cut-off for the $CFT_{1+1}$.
Observe from the above expression that the contributions from various geodesics in eq.(\ref{ENCFT3}) cancel exactly in the bipartite limit $L\to\infty$ ( $B\to A^c$), except twice the length of the geodesic anchored on the subsystem-$A$. Hence, upon using the Brown-Hennaux formula $c=\frac{3R}{2G_{N}^{(3)}}$ the above equation reduces to
\begin{equation}
{\cal E} =\frac{c}{2}\ln\big[\frac{l}{a}\big]=\frac{3}{2}S_A ,\label{hcfpu}
\end{equation}
Quite interestingly this result precisely matches with the universal result obtained by Calabrese et al.  \cite{Calabrese:2012ew, Calabrese:2012nk} given in eq.(\ref{ENCFTzero}) through the replica technique for the vacuum state of a $CFT_{1+1}$. Note that our result is in conformity with quantum information theory where the 
the entanglement negativity for a pure state is given by the Renyi entropy of order half.

\subsection{ Euclidean BTZ black hole}

Let us now consider a finite temperature mixed state of a holographic $CFT_{1+1}$ which is dual to a bulk  Euclidean BTZ black hole whose metric is given by
\begin{equation}
 ds^{2}= \frac{(r^2-r_h^2)}{R^2}d\tau_{E}^2+ \frac{R^2}{(r^2-r_h^2)} dr^2 +r^2 d\phi^2,\label{BTZm}
\end{equation}   
here, $\tau_{E}$ is the compactified Euclidean time $(\tau_{E}\sim \tau_E+\frac{2\pi R}{r_h})$. Note that the coordinate $\phi$ is identified with $(\phi+2\pi)$ for the BTZ black hole and is uncompactified for the BTZ black string case as $\phi=\frac{x}{R}$. Under the co-ordinate transformation $r= r_h \cosh\rho, \tau_E=\frac{R^2}{r_h}\theta, \ \phi=\frac{R}{r_h}t $ the metric in (\ref{BTZm}) becomes
 \begin{equation}
  ds^2= R^2(d\rho^2+ \cosh^2{\rho} dt^2+ \sinh^2\rho d\theta^2). \label{EBTZm}
 \end{equation}
 The length of a space like geodesic $ {\cal L}_{\gamma}$  anchored on the subsystem $\gamma$ may then be determined using the above transformation as described in \cite{Nishioka:2009un}
\begin{equation}
 {\cal L}_{\gamma}=2R\ln\bigg[\frac{\beta}{\pi a}\sinh[\frac{\pi l_{\gamma}}{\beta}]\bigg],\label{lij}
\end{equation}
 where, $a$ is the UV cut-off for the $CFT_{1+1}$. The parameter $l_{\gamma}$ represent the length of the subsystem $(\gamma)$ in the holographic $CFT_{1+1}$. 
 The above general expression for the geodesic length ${\cal L}_{\gamma}$ in eq.(\ref{lij}) may then be utilized to obtain the corresponding lengths of various geodesics anchored on the appropriate subsystems $\gamma=\{A,B_1,B_2,A\cup B_1, A\cup B_2\}$ as depicted in the fig(\ref{fig:subsystem}). Using these expressions in eq.(\ref{ENCFT4}), the holographic entanglement negativity for the finite temperature mixed state in question may then be obtained as follows 
\begin{eqnarray}
 {\cal E}&=&\frac{3R}{4G_{N}^{(3)}}\bigg[\ln\bigg\{\frac{\beta}{\pi a}\sinh(\frac{\pi\ell}{\beta})\bigg\}-\frac{\pi \ell}{\beta}\bigg],\label{HENCFTfinite1}\\
 &=&\frac{3}{2}[S_{A}-S_{A}^{th}].\label{HENCFTfinite}
\end{eqnarray}
Remarkably the above result matches exactly with the universal result given by eq.(\ref{ENCFTlc}) for the entanglement negativity of a finite temperature mixed state in a $CFT_{1+1}$ obtained through the replica technique \cite{Calabrese:2014yza}, in the large central charge limit. Note that we have once again used the Brown-Hennaux formula $c=\frac{3R}{2G_{N}^{(3)}}$ to arrive from eq.(\ref{HENCFTfinite1}) to eq.(\ref{HENCFTfinite}). The expression in eq.(\ref{HENCFTfinite}) clearly suggests that the holographic entanglement negativity captures the distillable quantum entanglement for the finite temperature bipartite mixed state
of a $CFT_{1+1}$ through the elimination of the thermal contributions.

The above observations clearly suggests a higher dimensional extension of our holographic entanglement negativity
conjecture in the $AdS_3/CFT_2$ scenario to a generic \linebreak $AdS_{d+1}/CFT_{d}$ framework. To this end we first consider an extended bipartite system $(A\cup A^c)$ described by the holographic $CFT_d$. We then consider two other subsystems $B_1$ and $B_2$ within $A^c$, on either side of the subsystem $A$ such that $B=(B_1\cup B_2)$. If ${\cal A}_{A}$, ${\cal A}_{B_1}$ and ${\cal A}_{B_2}$ denote the areas of co-dimension two bulk $AdS_{d+1}$ static minimal surfaces which are anchored on the subsystems $A,B_1$ and $B_2$, then the holographic entanglement negativity for the bipartite system
$A\cup A^c$ described by the holographic $CFT_d$ is given as
\begin{equation}
 {\cal E}=\lim_{B \to A^c}\frac{3}{16G_{N}^{(d+1)}}\bigg[(2 {\cal A}_{A}+{\cal A}_{B_1}+{\cal A}_{B_2}-{\cal A}_{A\cup B_1}-{ \cal A}_{A\cup B_2})\bigg]. \label{ENCFT5}
\end{equation}
Here $G_{N}^{(d+1)}$ is the $(d+1)$-dimensional Newton constant and the limit $(B \to A^{c} )$ in eq.(\ref{ENCFT5}) corresponds to the bipartite limit in which subsystems $B_1$ and $B_2$ are extended infinitely such that the subsystem $B=(B_1 \cup B_2)$ becomes the rest of the system $A^c$. Once again upon using the Ryu-Takayanagi conjecture given in  eq.(\ref{EEI}), the expression in eq.(\ref{ENCFT5}) reduces to the following form
\begin{equation}
 {\cal E}=\lim_{B \to A^c}\frac{3}{4}\bigg[2 S_{A}+S_{B_1}+S_{B_2}-S_{A\cup B_1}-S_{A\cup B_2}\bigg]. \label{ENCFTSAd}
\end{equation}
The above expression for the holographic entanglement negativity may be re-expressed as the sum of holographic mutual informations ${\cal I}(A,B_i)$ between appropriate subsystems as follows
 \begin{eqnarray}
 {\cal E}=\lim_{B\to A^{c}} \frac{3}{4}\big[{\cal I}(A,B_1)+{\cal I}(A,B_2)\big]\label{ENMII}
  \end{eqnarray}
where, the holographic mutual information ${\cal I}(A,B_i)$ ($i=1,2$)  are given as follows
 \begin{eqnarray}
  {\cal I}(A,B_i)&=&S_{A}+S_{B_i}-S_{A\cup B_i},\nonumber\\
&=&\frac{1}{4G_{N}^{(d+1)}}({\cal A}_{A}+{ \cal A}_{B_i}-{\cal A}_{A\cup B_i}).
 \end{eqnarray}

We should mention here that in \cite{Chaturvedi:2016rft} we have employed our conjecture to 
compute the entanglement negativity for specific bipartite quantum states of a holographic $CFT_d$. This includes
the bipartite pure state described by the vacuum of the $CFT_d$ dual to the bulk pure $AdS_{d+1}$ space time and the finite temperature mixed state dual to the bulk $AdS_{d+1}$-Schwarzschild black hole. We were able to demonstrate
that for both the cases the holographic entanglement negativity following from our conjecture has identical forms as the universal $CFT_{1+1}$ results given in eq. (\ref {hcfpu}) and eq. (\ref {HENCFTfinite}) respectively. Although this serves as a fairly strong consistency check we should emphasize that a proof for the higher dimensional extension of our conjecture along the lines of \cite{Lewkowycz:2013nqa} remains a significant open issue that needs to be addressed.

\section{Summary and Conclusion}
 To summarize we have advanced a holographic conjecture for the entanglement negativity of  bipartite systems described by holographic $CFT_{1+1}$s in the $AdS_3/CFT_2$ scenario. As a first consistency check
of our conjecture we have computed the holographic entanglement negativity for specific bipartite quantum states of a $CFT_{1+1}$ utilizing our conjecture. These were exemplified by the pure state of the $CFT_{1+1}$ vacuum dual to the pure $AdS_3$ space time and the finite temperature mixed state dual to the 
Euclidean BTZ black hole. Remarkably for both the pure and the mixed states, the holographic entanglement negativity computed from our conjecture exactly reproduces the corresponding replica technique results for the $CFT_{1+1}$ in the large central charge limit. For the pure state of the $CFT_{1+1}$ vacuum, our conjecture leads to the expected result from quantum information theory that the entanglement negativity is equal to the Renyi entropy of order half. Furthermore for the finite temperature bipartite mixed state of the $CFT_{1+1}$, it is observed that the holographic entanglement negativity captures the distillable entanglement through the elimination of the thermal contributions. Significantly the above results exactly reproduce the corresponding universal part of the replica technique results for the entanglement negativity of the holographic $CFT_{1+1}$. As mentioned earlier despite the strong consistency check exemplified by the above results and the clear substantiation through the large central charge analysis in \cite {Malvimat:2017yaj}, a proof along the lines of \cite {Faulkner:2013yia} for our conjecture remains a critical open issue.

Our holographic entanglement negativity conjecture in the $AdS_3/CFT_2$ context naturally indicates a
higher dimensional extension which we have briefly alluded to in our article. The application of the higher dimensional extension to specific examples have led to interesting results in \cite {Chaturvedi:2016rft} which may serve as a first consistency check. However we mention here that in the absence of independent explicit computations for the $CFT_d$ our negativity conjecture requires a proof along the lines of \cite {Lewkowycz:2013nqa}. As mentioned earlier this remains a significant open issue for the future.

Quite clearly our proposed holographic entanglement negativity conjecture in the $AdS_3/CFT_2$ context
is expected to have significant implications for the characterization of mixed state entanglement in diverse condensed matter phenomena such as quantum criticality and topological phases amongst others. Furthermore
a covariant version of holographic entanglement negativity in the $AdS_3/CFT_2$ scenario, may be useful to address issues of quantum gravity such as thermalization, black hole formation and collapse phenomena, information loss paradox and the related firewall problem. The possible higher dimensional extension, if strongly substantiated, may also lead to interesting insights into diverse physical phenomena. These constitute fascinating open issues for future investigations.

\section{Acknowledgement}
The authors would like to thank Ashoke Sen for crucial suggestions and K. S. Narain for interesting discussions. We would also like to thank Sayantani Bhattacharyya and Saikat Ghosh for extremely useful discussions and significant insights. The work of Pankaj Chaturvedi is supported by  Grant No. 09/092(0846)/2012-EMR-I, from the Council of Scientific and Industrial Research (CSIR), India.

\bibliographystyle{utphys} 

\bibliography{H_ENbib} 
\end{document}